\documentclass[amsmath,amssymb,superscriptaddress,floatfix,reprint,onecolumn]{revtex4}
\usepackage{bm}         
\usepackage{amssymb}
\usepackage{amsmath}
\usepackage{graphicx}
\usepackage{epsfig}
\usepackage{color}
\usepackage{ulem}
\usepackage{SIunits}

\usepackage[dvips,colorlinks]{hyperref}
\usepackage[all]{hypcap}
\usepackage{prettyref}

\def\equationautorefname~#1\null{Equation~(#1)\null}

\hypersetup{pdftitle={},pdfauthor={},
            pdfkeywords={},
            bookmarksnumbered,pdfstartview={FitH},urlcolor=red,linkcolor=black,citecolor=black}

\makeatletter
\def\Yb{\@ifnextchar.{\@YbS}{\Y@b}}
\def\@YbS{YbRh$_2$Si$_2$}
\def\Y@b{YbRh$_2$Si$_2$\hspace{1ex}}
\makeatother
\def\rf{\ensuremath{_{\rm rf}}}
\def\etal{{\sl et~al. }}

\newcommand{\norm}[1]{\ensuremath{\left|#1\right|}}

\addunit{\Gauss}{Gs}

\begin{document}

\title{Optimization of Coplanar Waveguide Resonators for ESR Studies on Metals}
\author{Conrad Clauss, Martin Dressel, and Marc Scheffler}
\affiliation{1. Physikalisches Institut, Universität Stuttgart, Pfaffenwaldring 57, 70550 Germany}
\email{scheffl@pi1.physik.uni-stuttgart.de}
\begin{abstract}
\noindent{\bf Abstract}\hspace{1.5em} We present simulations and analytic calculations of the electromagnetic microwave fields of coplanar waveguide (CPW) resonators in the vicinity of highly conducting metallic samples. The CPW structures are designed with the aim of investigating electron spin resonance (ESR) in metallic heavy-fermion systems, in particular YbRh$_2$Si$_2$, close to the quantum critical point. Utilizing CPW resonators for ESR experiments allows for studies at $\milli\kelvin$ temperatures and a wide range of freely selectable frequencies. It is therefore of great interest to evaluate the performance of resonant CPW structures with nearby metallic samples. Here we study the microwave fields at the sample surface as a function of sample distance from the waveguide structure and analyze the implications of the sample on the performance of the resonator. The simulation results reveal an optimum sample distance for which the microwave magnetic fields at the sample are maximized and thus best suited for ESR studies.
\end{abstract}

\maketitle

\fontsize{11pt}{1.2\baselineskip}\selectfont
\section{Introduction}
Quantum criticality in heavy-fermion metals is a major research topic in solid state physics \cite{Loehneysen07,Gegenwart08,Si10}. The main concept is that a magnetic phase transition is suppressed continuously with an external tuning parameter, which changes the relative strength of RKKY interaction (favoring magnetic order) and Kondo interaction (favoring a non-magnetic ground state). Here the $f$ electrons of Ce or Yb atoms in the crystal structure play a crucial role, and therefore studying the spin dynamics of local magnetic moments and/or heavy quasiparticles is of great interest \cite{Scheffler2013}. A well-established example for heavy-fermion quantum criticality is YbRh$_2$Si$_2$, where a small magnetic field of 60~mT applied within the tetragonal crystal plane suffices to suppress the antiferromagnetic order that is present below 70~mK at zero magnetic field \cite{Gegenwart08,Si10}.
YbRh$_2$Si$_2$ exhibits pronounced ESR at low temperatures \cite{Sichelschmidt03}. In conventional ESR experiments, the resonance condition $hf = g \mu_B H$ (with $h$ Planck's constant, $f$ microwave frequency, $g$ ESR $g$-factor, $\mu_B$ Bohr magneton, and $H$ applied magnetic field) is fulfilled by setting a fixed ESR frequency $f$ and then sweeping the magnetic field $H$ until resonance is achieved. For YbRh$_2$Si$_2$, sweeping the external magnetic field also means moving in the phase diagram of the material, and therefore it is desirable to study YbRh$_2$Si$_2$ at different ESR fields, which requires different ESR frequencies. While numerous commercial as well as high-frequency ESR setups have already been used to study YbRh$_2$Si$_2$ \cite{Sichelschmidt03,Wykhoff07b,Schaufuss09,Duque09,Sichelschmidt10b}, they could not access the most interesting regimes of the phase diagram due to two fundamental limitations. Firstly, conventional ESR techniques only work down to $^4$He temperatures, or with extra effort down to $^3$He temperatures \cite{Sichelschmidt10b}, and thus cannot come close to the low temperatures needed to reach the antiferromagnetic and Fermi-liquid regimes \cite{Gegenwart08}. Secondly, they are limited in their choice of ESR frequency, and therefore cannot be adjusted to the frequencies required for particular magnetic fields of interest.
Concerning frequency-tunable or broadband ESR, several new techniques have been developed recently \cite{Boero03,Narkowicz05,Jang08,Schlegel10,Harward11,Clauss13}. An approach that is particularly suited to overcome the limitations discussed above are planar microwave resonators \cite{Scheffler2013}, which are compact enough for operation in a dilution refrigerator at mK temperatures and provide great freedom of choice regarding the resonance frequency due to the simple possibility of designing and implementing the resonator according to the desired characteristics.\\
In the present work, we show how ESR measurements on a metallic sample using coplanar resonators can be optimized, in particular by changing the distance between sample and resonator chip. Here we have the special case of YbRh$_2$Si$_2$ in mind, but there are more heavy-fermion materials where such an experiment is of interest \cite{Friedemann2009,Holanda2011}. Furthermore, planar microwave resonators similar to the one discussed here are also used in quantum information science or for spectroscopy on the conductivity of superconductors and metals \cite{Goeppl2008,Scheffler2012,Hafner2014}.
\section{Resonator Model and Parameters}
\begin{figure}[htb]
	\centering
	\includegraphics[width=.9\textwidth]{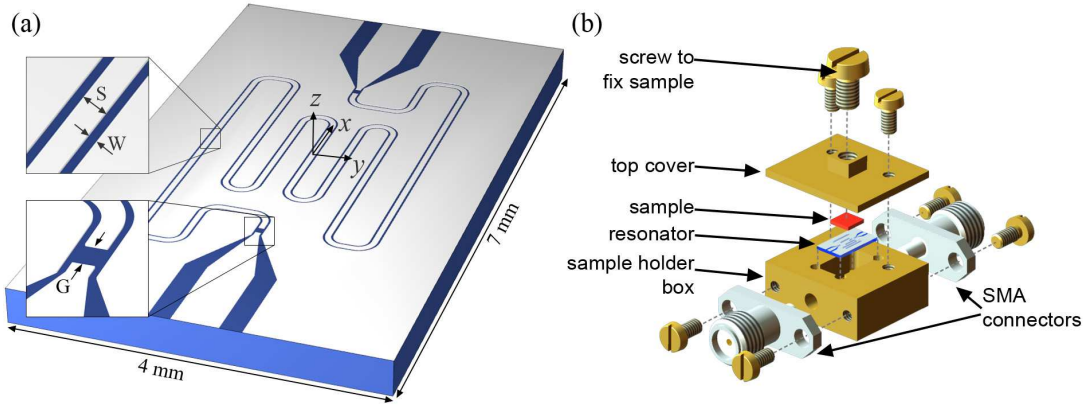}
	\caption{\fontsize{11pt}{1.2\baselineskip}\selectfont
(a) Sketch of the layout of the coplanar resonator. The metallic film is shown in light gray on top of a blue colored substrate. The relevant geometrical parameters $S$, $W$ and $G$ are shown in the zoom-in view. (b) Exploded drawing of the experimental realization including the waveguide resonator, the sample, and microwave connectors.}
	\label{fig:Fig1}
\end{figure}

The resonator under study is shown in \autoref{fig:Fig1}~(a) and consists of a structured superconducting film on top of a dielectric substrate (sapphire). The structure itself is comprised of a signal carrying center conductor flanked by two ground planes. For a given relative permittivity $\varepsilon_r$ of the substrate, the waveguide impedance, which has to be matched to the $50~\ohm$ impedance of the microwave equipment, is mainly defined by the width $S$ of the center conductor and the distance $W$ between center conductor and ground planes \cite{Simons01} (here $S=60~\micro\meter$ and $W=25~\micro\meter$ for a sapphire substrate with $\varepsilon_r=10$). 
%
%
The boundary conditions -- for instance distances to lateral or vertical conductive walls -- have also to be taken into account once they get comparable with the structural dimensions of the waveguide. Since the microwave electric and magnetic fields extend both into the substrate and the space above the structure (here modeled as air), the waveguide properties can be expressed in terms of an effective dielectric constant $\varepsilon_{\rm eff}$ which is a function of all structural characteristics, including the boundary conditions \cite{Simons01}. To create a coplanar waveguide resonator, it suffices to integrate two discontinuities at a distance $\ell$ into the center conductor that allow for partial reflection of the microwave. A common way is inserting two gaps of size $G$ that capacitively couple the adjacent center conductor strips. The size of $G$ determines the coupling strength (here $G$ was chosen to $100~\micro\meter$). The resonator frequencies are then given by $f_n=(n+1)c/(2\ell\sqrt{\varepsilon_{\rm eff}})$ (with $n=0,1,2,\ldots$ and $c$ the vacuum speed of light).\\
The center conductor of the structure depicted in \autoref{fig:Fig1}~(a) is widened towards the input and output feeds to accommodate for the microwave connectors. For the actual experiment, the waveguide resonator is put into a brass sample holder box, connected to coaxial lines and the sample is affixed to a bolt in the lid of the sample holder box to adjust for different heights (cf. \autoref{fig:Fig1}~(b)).\\
%
\begin{figure}[htb]
	\centering
	\includegraphics[width=.85\textwidth]{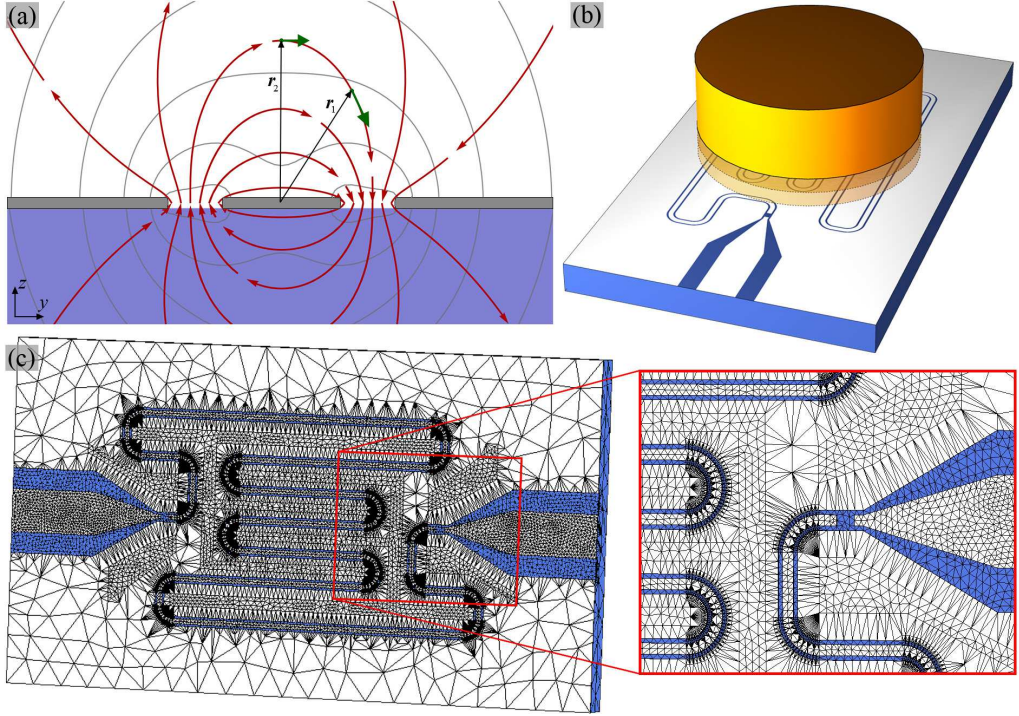}
	\caption{\fontsize{11pt}{1.2\baselineskip}\selectfont
(a) Magnetic fields in the cross section of a coplanar waveguide. Displayed are the field lines (arrows) and the relative field strength (contour lines, drawn in intervals of $2^n$). Two points are marked ($\boldsymbol r_{1,2}$) to exemplarily illustrate the position dependence of the polarization vector. (b) Model to simulate the influence of a conducting sample close to the resonator. The sample is represented by a cylindrical shape with a variable distance to the waveguide structure. (c) Illustration of the mesh cells of the model. To reliably obtain the local fields, the structure needs to be densely meshed close to the waveguide (zoom view for more detail).}
	\label{fig:Fig2}
\end{figure}
%
\\
The structure studied in this work was designed for a fundamental frequency of $f_0=2.68~\giga\hertz$ in the absence of any sample material. In such a case, the microwave magnetic field ($H\rf$) shows a profile as illustrated in \autoref{fig:Fig2}~(a) (calculated according to \cite{Simons82}). The polarization of $H\rf$ is a strongly position-dependent superposition of linear polarization along $y$ and along $z$-direction. The sample was modeled as a cylindrical metallic object with a diameter of $3~\milli\meter$ and a resistivity of $0.6~\micro\ohm\;\centi\meter$ (low temperature resistivity of \Yb \cite{Gegenwart08b}) placed at a distance $d$ above the center of the waveguide resonator structure. For simplicity, the waveguide structure itself was modeled as perfect electric conductor. To obtain the electromagnetic fields, the volume is divided into mesh cells (see \autoref{fig:Fig2}~(c)), and the $S$-parameters (scattering parameters) as well as the field distributions are acquired using CST Microwave Studio$^\circledR$.
\section{Results}
\begin{figure}[t]
	\centering
	\includegraphics[width=.92\textwidth]{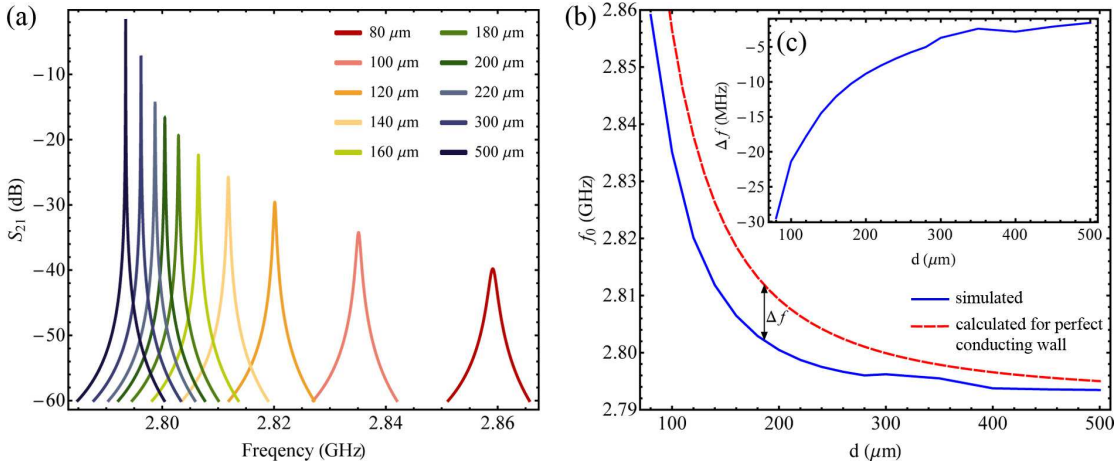}
	\caption{\fontsize{11pt}{1.2\baselineskip}\selectfont
(a) Forward transmission coefficient $S_{21}$ as a function of frequency for a selection of sample distances $d$. (b) Sample distance dependence of the resonance frequency of the simulated results (solid blue) and analytic results (dashed red) as a consequence of changed boundary conditions. (c) Resonance frequency shift due to dissipative losses in the sample.}
	\label{fig:Fig3}
\end{figure}

\begin{figure}[htb]
	\centering
	\includegraphics[width=.98\textwidth]{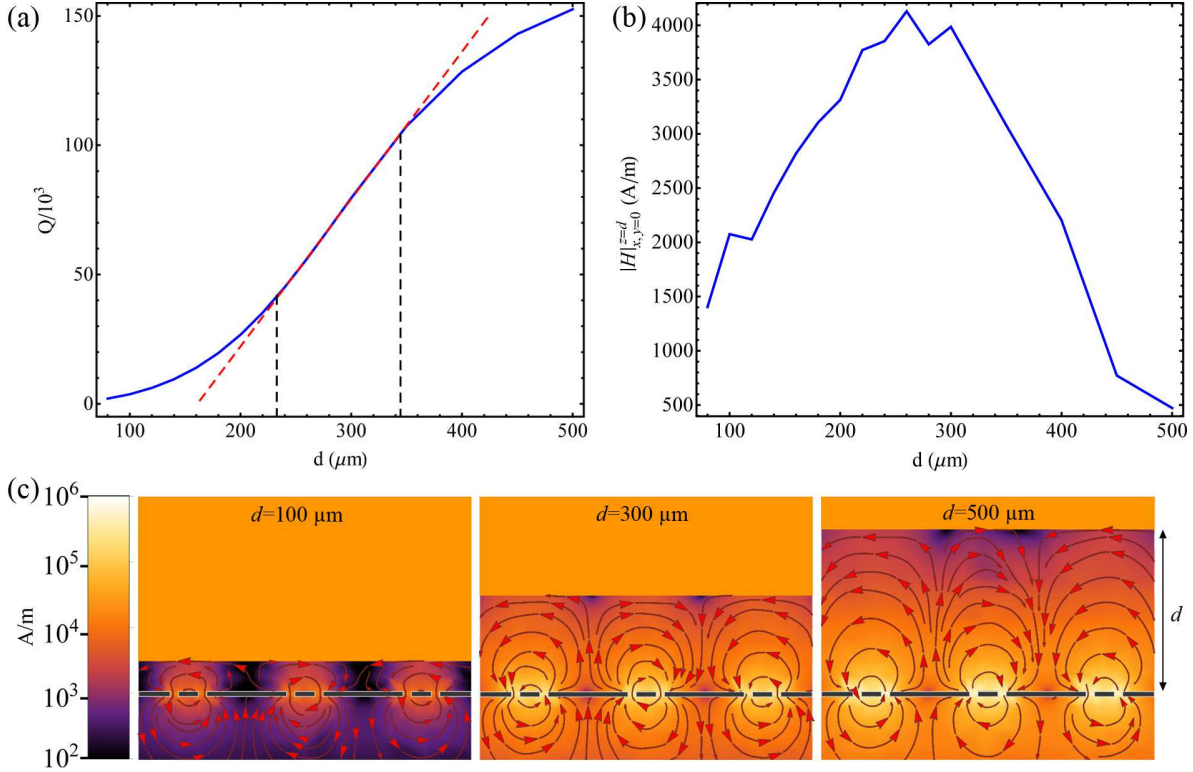}
	\caption{\fontsize{11pt}{1.2\baselineskip}\selectfont
(a) Quality factor $Q$ of the resonator as function of sample distance. The linear regime approximately marks the region of strong magnetic fields at the sample. (b) Absolute magnetic field strength directly at the sample/air interface -- the graph shows a clear maximum around $d\approx 270~\micro\meter$. (c) Cross section through the model in the $yz$-plane displaying the field distribution in the space between sample and resonant structure.}
	\label{fig:Fig4}
\end{figure}

In the absence of any sample, the metallic cover is located $3~\milli\meter$ above the resonator structure. For such a scenario the simulated forward transmission ($S_{21}$ parameter) shows a clear Lorentzian shaped resonance centered around $2.792~\giga\hertz$. In this work, the structure was simulated for sample distances $d$ ranging from 80 to $500~\micro\meter$. A selection of $S_{21}$ transmission parameter curves is plotted in \autoref{fig:Fig3}~(a). As a function of decreasing sample distance, three effects become clearly notable -- (i) the resonance frequency is shifted towards higher frequencies; (ii) the height of the resonances strongly decreases, and (iii) the width of the resonance curves increases substantially.%
%
%
While the two latter observations can be explained in terms of supplemental ohmic losses (see below) the first point is a superposition of two different effects. To explain this behavior, the obtained frequencies are plotted as a function of sample distance in \autoref{fig:Fig3}~(b) (solid blue line). In addition, the resonance frequency was calculated for a structure with a perfect conducting electric wall placed at distance $d$ (dashed red line) via $f_0=c/(2\ell\sqrt{\varepsilon_{\rm eff}(S,W,d)})$. Here, $\ell$ was chosen such that both frequencies coincide for $d=3~\milli\meter$ at $f_0=2.792$. The result shows the same global behavior as the simulated resonance frequencies but with a distance dependent offset. Therefore, the increase of frequency is a result of changed boundary conditions and only the offset describes the frequency shift due to the ohmic losses caused by the sample (shown in \autoref{fig:Fig3}~(c)).\\
A commonly used quantity to describe the performance of resonators is the quality factor $Q$ which directly probes the losses in the complete system. It is defined as the ratio between energy stored in the resonator and energy dissipated per cycle. Experimentally, it can be obtained by $Q=f_0/\Delta f_0$ (with $\Delta f_0$: FWHM of the resonance curve). The $d$-dependent quality factor is plotted in \autoref{fig:Fig4}~(a). At large sample distances, $Q$ is on the order of $10^5$ and decreases strongly over almost two orders of magnitude with decreasing $d$. This reduction of $Q$ is purely due to the finite resistivity of the sample material, since $Q$ stays constant for perfect conducting samples (not shown), and illustrates the strong dependence of the resonator performance on the presence of any loss channels.
%
%
\\
From the viewpoint of applicability concerning ESR studies the absolute microwave magnetic field at the sample position is of the paramount interest since the ESR line intensity scales with $H\rf^2$ \cite{Abragam-Bleaney70}. Figures \ref{fig:Fig4}~(b) and (c) illustrate how the magnetic fields depend on the sample distance for an input power of $1~\watt$. Panel (b) shows the absolute value of the rf field amplitude $\left|H\rf\right|$ directly at the sample surface above the center of the waveguide structure at $(0,0,d)$ and panel (c) shows the field distribution in a cross section of the model structure in the $yz$-plane ($x=0$, $\norm y\leqslant 500~\micro\meter$, $-200\leqslant z\leqslant 600~\micro\meter$, see \autoref{fig:Fig1} (a) for definition of coordinates). For large distances, the system gets only slightly perturbed by the lossy sample. The quality factor remains high since the electric and magnetic fields at the sample surface are very small. At very small sample distances, the damping of the resistive sample material is dominant and the quality factor is strongly suppressed. As a result, the magnetic fields at the sample-air interface are again small as most of the microwave power is dissipated in the sample. In an intermediate region, for this particular structure ranging from $\sim 220$ to $\sim 320~\micro\meter$, the magnetic fields at the sample surface show a clear maximum and $Q$ scales linearly with sample distance. For those values of $d$ the resonator efficiency parameter \cite{Hyde89}
$$\Lambda=\frac{H\rf^{\rm max}(z=d)}{\sqrt{P_{\rm in}}}$$
amounts to $\Lambda>45~\Gauss\sqrt\watt$ (with incident power $P_{\rm in}=1~\watt$ and $1~\Gauss\widehat{=}1000/(4\pi)~\ampere\meter^{-1}$). This parameter gives an impression of how well microwave power is transformed into microwave magnetic field at the sample and thus provides the possibility to compare different ESR resonators. The values obtained in this work exceed those of standard ESR cavities (X-band) by a factor $>30$ \cite{Hyde89,Anderson02,Hyde02} demonstrating very good performance of CPW resonators for ESR purposes.\\
Another important aspect which can be seen in \autoref{fig:Fig4}~(c) is that for the intermediate distances the polarization of the microwave field at the sample surface is almost exclusively along $y$-direction, allowing for a clear assignment of the ESR signal in case of anisotropic metallic samples.\\
Since the spatial confinement of the microwave fields depends on the geometrical parameters of the waveguide structure, i.e. the center conductor width $S$ and the center conductor-ground plane separation $W$, the optimum region of $d$ varies for different geometries. It is therefore worthwhile to record the quality factor as a function of distance and to choose a distance in the linear regime for the ESR measurement.\\
In addition, the microwave input power has to be low enough to avoid heating of the sample due to induced currents. The surface current density is identical to the fields given in \autoref{fig:Fig4}~(b). With only hundreds of $\micro\watt$ of cooling power at $\milli\kelvin$ temperatures the microwave power has to be reduced by several orders of magnitude to avoid this unintended heating.
\section{Summary}
Using electromagnetic simulations, we have shown that CPW resonators are well suited for ESR on metallic, highly conducting samples such as quantum-critical \Yb. As a function of sample distance the changes of the resonance frequency are small whereas the quality factor changes drastically.
It was demonstrated that for a certain range of sample distances large efficiency parameters $\Lambda$ are possible and that this range can be determined by a distance-dependent measurement of the quality factor.\\[+\baselineskip]
We thank Helga Kumrić for helpful support with the simulations. We acknowledge financial support by the Deutsche Forschungsgemeinschaft (DFG) including SFB/TRR21.



\begin{thebibliography}{27}%
	\fontsize{11pt}{1.2\baselineskip}\selectfont
\bibitem{Loehneysen07} v. Löhneysen H, Rosch A, Vojta M, and Wölfle P 2007, {\it Rev. Mod. Phys.} {\bf 79} 1015
\bibitem{Gegenwart08} Gegenwart P, Si Q, and Steglich F 2008, {\it Nat. Phys.} {\bf 4} 186
\bibitem{Si10} Si Q and Steglich F 2010, {\it Science} {\bf 329} 1161
\bibitem{Scheffler2013}Scheffler M \etal 2013, {\it Phys. Status Solidi} B \textbf{250} 439
\bibitem{Sichelschmidt03} Sichelschmidt J, Ivanshin V A, Ferstl J, Geibel C, and Steglich F 2003, {\it Phys. Rev. Lett.} {\bf 91} 156401
%
%
\bibitem{Wykhoff07b} Wykhoff J, Sichelschmidt J, Lapertot G, Knebel G, Flouquet J, Fazlishanov I I, Krug von Nidda H-A, Krellner C, Geibel C, and Steglich F 2007, {\it Sci. Technol. Adv. Mater.} {\bf 8} 389
\bibitem{Schaufuss09} Schaufuss U, Kataev V, Zvyagin A A, Büchner B, Sichelschmidt J, Wykhoff J, Krellner C, Geibel C, and Steglich F 2009, {\it Phys. Rev. Lett.} {\bf 102} 076405
\bibitem{Duque09} Duque J G S \etal 2009, {\it Phys. Rev.} B {\bf 79} 035122
%
%
\bibitem{Sichelschmidt10b} Sichelschmidt J \etal 2010, {\it Phys. Status Solidi} B {\bf 247} 747
%
\bibitem{Boero03} Boero G, Bouterfas M, Massin C, Vincent F, Besse P-A, Popovic R S, and Schweiger A 2003, {\it Rev. Sci. Instrum.} {\bf 74} 4794
\bibitem{Narkowicz05} Narkowicz R, Suter D, and Stonies R 2005, {\it J. Magn. Reson.} {\bf 175} 275
%
\bibitem{Jang08} Jang Z H, Suh B J, Corti M, Cattaneo L, Hajny D, Borsa F, and Luban M 2008, {\it Rev. Sci. Instrum.} {\bf 79} 046101
\bibitem{Schlegel10} Schlegel C, Dressel M, and van Slageren J 2010, {\it Rev. Sci. Instrum.} {\bf 81} 093901
\bibitem{Harward11} Harward I, O’Keevan T, Hutchison A, Zagorodnii V and Celinski Z 2011, {\it Rev. Sci. Instrum.} {\bf 82} 095115
\bibitem{Clauss13} Clauss C, Bothner D, Koelle D, Kleiner R, Bogani L, Scheffler M, and Dressel M 2013, {\it Appl. Phys. Lett.} {\bf 102} 162601
\bibitem{Friedemann2009} Friedemann S, Westerkamp T, Brando M, Oeschler N, Wirth S, Gegenwart P,  Krellner C, Geibel C, and Steglich F 2009, {\it Nat. Phys.} {\bf 5} 465
\bibitem{Holanda2011} Holanda LM, Vargas JM, Iwamoto W, Rettori C, Nakatsuji S, Kuga K, Fisk Z, Oseroff SB, and Pagliuso PG 2011, {\it Phys. Rev. Lett.} {\bf 107} 026402
\bibitem{Goeppl2008} Göppl M, Fragner A, Baur M, Bianchetti R, Filipp S, Fink J M, Leek P J, Puebla G, Steffen L, and Wallraff A 2008, {\it J. Appl. Phys.} {\bf 104} 113904
\bibitem{Scheffler2012}Scheffler M, Fella C, and Dressel M 2012, {\it J. Phys.: Conf. Series} {\bf 400} 052031
\bibitem{Hafner2014}Hafner D, Dressel M, and Scheffler M 2014, {\it Rev. Sci. Instrum.} {\bf 85} 014702
\bibitem{Simons01} Simons R N 2001, {\it Coplanar Waveguide Circuits, Components, and Systems} (New York: Wiley-Interscience)
\bibitem{Simons82} Simons R N and Arora R K 1982, {\it IEEE Trans. Microw. Theory Techn.} {\bf 30} 1094
\bibitem{Gegenwart08b} Gegenwart B, Westerkamp T, Krellner C, Brando M, Tokiwa Y, Geibel C, and Steglich F 2008, {\it Physica} B {\bf 403} 1184
\bibitem{Abragam-Bleaney70} Abragam A and Bleaney B 1970, {\it Electron Paramagnetic Resonance of Transition Ions} (Oxford: Oxford University Press)
\bibitem{Hyde89} Hyde J S and Froncisz W 1989, {\it Advanced EPR: Applications in Biology and Biochemistry}, edited by Hoff A J (Amsterdam: Elsevier)
\bibitem{Anderson02} Anderson J R, Mett R R, and Hyde J S 2002, {\it Rev. Sci. Instrum.} {\bf 73} 3027
\bibitem{Hyde02} Hyde J S, Mett R R, and Anderson J R 2002, {\it Rev. Sci. Instrum.} {\bf 73} 4003
%
\end{thebibliography}
\end{document}